\documentclass[conf]{IEEEtran}

\usepackage{graphicx}

\begin{document}

\title{Short Paper: Cells in the Internet of Things}

\maketitle

\begin{abstract}
The Internet of Things combines various earlier areas of research. As a result, research on the subject is still organized around these pre-existing areas: distributed computing with services and objects, networks (usually combining 6lowpan with Zigbee etc. for the last-hop), artificial intelligence and semantic web, and human-computer interaction. We are yet to create a unified model that covers all these perspectives - domain, device, service, agent, etc. In this paper, we propose the concept of ”cells” as units of structure and context in the Internet of things. This allows us to have a unified vocabulary to refer to single entities (whether dumb ”motes”, intelligent ”spimes”, or virtual ”services”), ”intranets” of things, and finally the complete Internet of things. We also mention how we can demarcate boundaries, and classify cells based on how they are accessed and controlled. Our aim is to present a unified model, to serve as the basis for formal modeling of security etc. in the Internet of Things.
\end{abstract}

\section{Introduction}

The Internet of Things has, in recent years, grown into a major area of research as well as commercial interest. The entire impact of the Internet has so far been caused by connecting together servers, sufficiently accessible personal computers (by which term we include desktops, laptops, and mobile phones), and finally people; pervasive intelligence at Internet scale can allow even more impactful applications - for example, Burrus\cite{burrus2014} mentions smart bridges that automatically detect when they are overloaded, and, communicating with smart stoplights and cars, redirect traffic away. The essential idea of the Internet of Things - to allow sensors and machines to connect and (in many cases) automatically configure themselves, in a heterogeneous, open architecture (with the Internet as a connectivity backbone) - is extremely powerful, because of its generality.  Existing research and plans for the actual build-out of the Internet of Things are more restricted in their scope. The primary Internet of Things initiatives\cite{atzori2010} are as follows.  
\begin{enumerate} 
\item EPC and uID: names for all objects.  
\item NFC, WSAN and RFID networks: linking objects to allow communication over the short range.  
\item IPSO, Internet : IP over anything.
\item ITU, CASAGRAS: enabling infrastructure for smart Things (spime) to sense and respond to context.
\item Semantic Web, Web of Things: Representing resources, communication, and computation.  
\end{enumerate}

These specific visions take one of three perspectives: they focus on the Internet, on the Things, or on the Data. (Morabito\cite{atzori2010} proposes that the third major component, rather than simply Data, is the Semantic Web.) Thus, while there are many different visions for the Internet of Things, there is not a single cohesive picture of its architecture (as components, connectors, and constraints) and function. In the absence of a universal system model, there are no well-defined roles. It is certainly possible to make out general principles (e.g. whether a particular design focuses on intelligence at the edges or intelligence in the pipes), but there is no standard method to classify what are colloquially called “Intranets of Things”, their loci of control, and how they are interconnected.

The second important “Tower of Babel” problem in the Internet of Things (which is ironic, given that the whole point of IoT is to solve a Babel problem itself and allow more communication!) is that approaches to the general problem - constructing an Internet of Things - are silo-ized along disciplinary lines. This is hardly surprising - the Internet of Things is a new area, which combines several earlier areas of research. We propose the following list:
\begin{enumerate}
\item Pervasive computing and ambient intelligence. (This topic includes smart homes, smart cars, smart highways etc.)
  \begin{enumerate}
  \item Physical computing: sensor (and actuator) networks.
  \item Context aware systems.
  \end{enumerate}
\item Networks, operating systems and middleware.
  \begin{enumerate}
  \item Mobile computing.
  \item Disruption-tolerant computing.
  \item Infrastructure abstraction - cloud computing etc.
  \end{enumerate}
\item Distributed computing and multi-agent systems.
  \begin{enumerate}
  \item Computing on graphs.
  \end{enumerate}
\end{enumerate}
There is a need to unify vocabulary and collect relevant results from each field.

Given the profusion of competing (though sometimes inter-composable) architectures, protocols, and middleware standards, the lack of standard vocabulary or models is a problem in making any precise statements about the Internet of Things. A particular issue is the lack of security: there is no one way to define the identity, responsibilities, and capabilities of a given entity, or to model threats and attacks. In this paper, we present such a general model for the Internet of Things, as a step towards enabling a clearer discussion of IoT principles and trends.

\section{Terms: Things and Roles}
In this section, we define the fundamental terms and concepts in our model of the Internet of Things.

The first term to define, with respect to “Internet of Things,” is thing. The most general definition of “thing” includes not only all devices and physical objects, but also people, logical entities (such as an airplane flight \cite{kelly2007}), and even the physical quantities (e.g. temperature) being measured or set by the devices. However, we are only concerned about things in the Internet of Things, so we propose a more precise definition. A thing is an entity that can send or receive messages.

A thing that simply acts as a sink for messages is not very interesting. These may be considered “dumb” things, as op- posed to “smart” things. However, the classification of things into “dumb” and “smart” is not clear. The original meaning of dumb was “incapable of speech”; however, it is now used for things which do not perform advanced information processing (as seen in a smartphone, smart house, or smart grid). There are at least two separate meanings being conveyed by “smart”, and we propose a third.
\begin{enumerate}
\item Ability to communicate.
\item Ability to sense and adapt to context.
\item Ability to be extended, by installing new applications etc.
\end{enumerate}

For clarity, we propose the term intelligent, to describe an entity that can detect and adapt to its context (say, by sending a message or actuating a physical event); conversant, to describe an entity which can respond to messages by sending replies; and extensible, to describe an entity which can act as a platform for new applications. We define a smart thing to be intelligent, conversant, and extensible.

It may further be noted that the words object, asset, and node are quite similar to thing. (It may be noted that the original meaning of the word “object” in object-oriented programming, as an “abstract computer” with separable interface and implementation, also involved the idea of an object being an end-point for messages \cite{kay1977}.) We propose that an object be defined as a thing that responds to a physical event (e.g. the temperature rose too high), or a message, with one or more physical events or messages. In other words, an object is an intelligent thing. We also propose the term actor to refer to objects which, on receiving a message, can not only take physical action or send messages, but can produce other objects \cite{hewitt1978}. We leave “node” as synonymous with “thing”. [In passing, we also suggest that although the Internet of Things certainly has roles for people (and social networks have been used as a component), it would perhaps be appropriate to not include them as “things”, and instead define the more general term “asset” to cover both people and things. Including people as things would introduce Sociology, Economics, and other studies of how people communicate and function. However, we will not pursue this line of inquiry further in this paper.]

Objects may be named, or not named. A named object can receive messages targeted toward it specifically.

An object, which corresponds to a distinct physical entity, is called a device. An object which does not correspond to a distinct physical entity, is called a service. Thus, a service is a coherent entity (with an end-point where messages can be sent), which provides functions while abstracting away the underlying infrastructure; the component objects that underlie the service and provide its capabilities, are not visible to the users of the service.

In this context, we will consider some common terms used in Internet of Things architecture: client, server, service, and broker. A common way to factor the Internet of Things\cite{roman2013} is using the client-server paradigm. In this view,
\begin{enumerate}
\item Client nodes are those where requests originate, typically machines with a human-computer interface where a user issues requests (or configures a program to do so).
\item Servers are named physical end hosts that respond to requests, again either by replying with a message (e.g. Web servers) or by taking a physical action (e.g. a “smart air conditioner”).
\item Services are named abstract entities, which also respond to requests, but hide the details of the physical implementation (ranging from Cloud Computing, to “smart grid” etc. infrastructure).
\item Brokers are nodes that match clients to servers or services, and perform any required translation (between types, protocols etc.) or orchestration (for example, “mashing up” existing services to provide the service a client wants). Broker nodes are essentially middleware services.
\end{enumerate}

For example, Lopez et al. \cite{roman2013} use an informal version of this taxonomy. Clients correspond to their Red nodes, servers to Blue, anonymous motes (infrastructure of a service) to Green, and brokers to Gray nodes. The authors develop a classification of Internet of Things architectures, based on the power of the brokers:
\begin{enumerate}
\item A broker acts as the central authority
\item A broker is one of several central authorities
\item A broker is a gateway
\item A broker is a simple free-floating component
\end{enumerate}

While our concept of cell-based architectures owes a great deal to this model, we do not feel that the client-server paradigm should be the main organizing principle of Internet of Things architecture. In the first place, there may well be confusion between the client and the server, or even broker, roles. Communication is not limited to data and requests for service; there may be contention for limited resources, like power, bandwidth, and even credit. (For a simple smart-home example, consider a smart fridge trying to chill the wine, and a smart bath trying to warm the water, in time for the owner to get home. The smart fuse warns them it is about to blow; both devices issue requests to each other, to ensure mutual exclusion. No component can clearly be identified as the client!) Secondly, the architecture may not be the same throughout the lifetime of a request. Consider fetching a simple webpage over HTTPS: two separate systems are needed - one for resource location (DNS), and one for authentication (X.509). The architecture of these systems may differ. Further, given that code, data and devices may all be mobile, the architecture may change during operation. For these reasons, we wish to develop a more general model for the Internet of Things, which allows client-server, peer-to-peer, and mixed component models, as well as centralized, distributed, and hybrid architectures.

\section{ Context and Cells}
The term context is very important in our work. Formally, “context” is the part of a discourse that is not part of the “text”, but modifies or explains its meaning. In Pervasive Intelligence, the term is used to describe the state of the situation surrounding an object; for example, objects sharing the same power supply share that context. We take the definition, “context is any information that can be used to characterise the situation of an entity” \cite{abowd1999}.

A cell is the unit within the boundary of which there exists a particular context. For example, a local area network with NAT is a cell, because IP addresses have a specific meaning within the cell (which is different from their semantics elsewhere). The range of an RFID card reader forms a cell around it. Any smart space (office, home, or car) is a cell.

A cell that is completely self-contained, and has no exchange of messages etc. with the rest of the IoT, may be called an “Intranet of Things”. Cells that are interconnected form a larger cell. In this model, a single device is a cell, a small network is a cell, and the entire Internet of Things is also a cell.

The natural question with regard to cells, is why we have two separate terms for objects and cells. As devices, services, and composite structures can be considered to be objects, why is it not sufficient to organize our architecture along object-oriented lines? (It may be argued that the IoT-A architecture \cite{neisse2014}, which considers the components of the Internet of Things to be Virtual Objects, Composite Virtual Objects, and Services, is an attempt to do roughly this.) The answer is that objects are defined by having an interface for communication - but communication is not the only organizing principle in the Internet of Things. We make note of at least three such principles, below.

\begin{enumerate}
\item \emph{Communication.} This is the traditional feature used to factor large systems such as the Internet. Communication cells are those which share a namespace (e.g. a LAN with local addresses), a medium of communication (e.g. the message bus in DDS), or a message format (wire protocol, API, etc.) In a system, the component that uses a specific protocol for some purpose can be considered a cell. For example, the Internet of Things consists of one large “backbone” cell running IP (usually IPv6, mIPv6, 6lowpan) and “leaf” cells using Bluetooth, Zigbee, Z-Wave etc.; these cells can be further factored into other, smaller cells.

As we consider namespaces to be part of communication, in our model identity and role are also part of this feature. The extent of the territory within which a name or its associated access rights are valid, is also a cell. (It may be noted that this item corresponds to “Who” and “What” in the five-W model of context [8]: When, Where, Who, What and Why.)

\item \emph{Physical Location.} A major factor in determining which cell a component belongs to, is its position. This feature is particularly significant in the Internet of Things because cyber-physical systems need a physical presence (sensor, actuator) at the position where an environmental value is sensed or changed; also, the pre-existing non-IP technology base (RFID, WSN etc.) is limited in terms of range. Position may be time-varying, as cells may be mobile, and have mobile components. For this reason, we
feel time is included as part of the physical location.(This item corresponds to “Where” and “When” in the five-W model - spatial and temporal cells.)

\item \emph{Power.} Things may be divided into active, i.e. power-using, and passive, e.g. RFID tags. Active things with an internal source of power (such as wireless sensors) are self-contained power cells; externally-powered active things belong to the power cell of the bus that delivers power. A “switching domain” for power is a cell. For passive things, the communication cell w.r.t. the physical medium of communication is also the power cell they belong to. (More generally, particular things require certain ambients to function - for example, light sensors require light. It may be possible to use these ambients as cell-defining features in particular contexts.)
\end{enumerate}

In the next section, we go into more detail about our architecture based on cells, and address the question of how cells are composed together.

\section{ Cell Structure, Contracts, and Connections}

The first question regarding a cell-based architecture, is what features are used to determine the boundaries of cells. In the previous section, we described three features of importance - communication (and control), physical location, and power. These features may also be used to classify cells. For this purpose, we introduce the concept of binding.

Binding refers to the contract enforced by a cell $C$. If component cells within cell $C$ have to ensure that they observe some property, then this property is a binding for $C$.
\begin{enumerate} 
\item \emph{Temporal} binding. Whether $C$ is synchronous or asynchronous, whether message delivery is real-time, and whether action on a message (reply, computation, physical action) is real-time. For example, most Internet of Things architectures are event-driven, and rarely synchronous; however, even some pub-sub architectures (such as DDS) make guarantees of very low latency.

\item \emph{Spatial} binding. Whether the cell is coupled to the physical location of its components (and whether the location of the cell itself, and the locations of components inside it, can vary with time). This includes questions of whether components have to be within a physical territory, within RFID reader range, and so on. An example is location-based access control.

It may be noted that enforcing spatial binding is itself a major challenge. A cell defined as a smart home, for example, runs into the immediate problem that the physical walls of the home and the territory reached by its wireless networks may not be the same. Consequently, it is possible that, say, a neighbor’s media system might take over speakers in the house. This is an even more serious concern in case of Body Area Networks.

\item \emph{Referential} binding. Whether the components of $C$ are named. Clearly, this can be further elaborated based on the point of view: internal or external. (A black-box cell may not allow any external object to see its components, but they may be named and addressable inside the cell itself.) By analogy with spatial binding, it may be possible to raise the question of whether names are time-dependent also; we take the simplest solution, and consider a cell a new entity every time it changes its name.
\end{enumerate}

[We considered a classification of cells by power, as passive/active and, if active, wired/wireless/battery powered. However, we could not create an example of a cell with only multiple passive components, that was also a meaningful composite entity. Hence we are not presenting Power binding.]

We now come to the question of the internal structure of a cell. It would be very appealing to keep the simple notion that ``everything is a cell''. Accordingly, we define every ``thing'' as a cell. However, this is not enough! In addition to components, a physical cell also has connectors, such as channels that carry power and messages. If these channels are not ``things'' by our definition, i.e. they are not capable of sending or receiving messages, can they be considered cells? (This is a hard question. It may be argued that what we define as a ``thing'' is actually a ``node'', and both nodes and edges should be called ``things'', but this would break the standard meaning of ``thing'' in IoT.) We take the position that connectors are not necessarily cells. A cell consists of a non-empty set of things, their connectors (if any), and the context that defines it.

Following Lopez \cite{roman2013}, we started with the following classes of structure for a cell.
\begin{enumerate}
\item Centralized. There is a single controller for the cell.
\item Collaborative. There are multiple controllers for the cell,
which are mutual peers.
\item Connected. A cell is composed of sub-cells, each with its
own controller, and the sub-cells are only connected by
connections between the controllers.
\item Distributed. A cell has no clear authority structure.
\end{enumerate}

\begin{figure}
\includegraphics[height=2in]{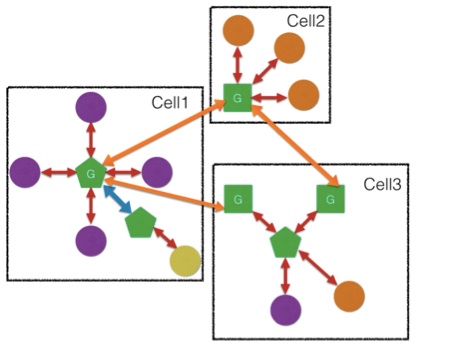}
\caption{ Architecture with some sample cells. Cell1 is Collaborative Single-Homed, Cell2 is Alliance Single-Homed, and Cell3 is Centralized Multi-Homed. The overall cell, having no controller or gateway, is Alliance Closed. (Legend: controllers are pentagons, gateways are marked ’G’.)}
\label{fig:1}
\end{figure}

However, this taxonomy has the disadvantage of no separation between the controller of a cell and its gateway. A controller is the locus of control, i.e. of definitive statements - for example, in a synchronous cell, the controller may be the master clock.  A gateway is the locus of some interface (communications, power, etc.) where the cell connects to other cells (and possibly to the world). For example, it might allow queries or commands to be sent to “all lights in No.$4$ Privet Drive” (a spatial cell).

Accordingly, we refine the above model. With respect to control, a cell is of the form:
\begin{enumerate}
\item \emph{Guardian.} There exists at least one controller. 
\begin{enumerate}
 \item Centralized. Single controller.
 \item Collaborative. Multiple controllers.
\end{enumerate}
\item \emph{Alliance.} There is no distinct controller.
\end{enumerate}
With respect to connectivity, a cell is of the form:
\begin{enumerate}
\item  \emph{Open.} The cell has at least one interface. 
\begin{enumerate} 
\item Single-homed. Single interface.
\item Multi-homed. Multiple interfaces.
\end{enumerate} 
\item \emph{Closed.} The cell has no interface.
\end{enumerate}

A sample with a few kinds of cell is shown in Figure \ref{fig:1}. 

\section{Concluding Remarks}

The Internet of Things is an old concept - it has been around since the 1970s as “Embedded Internet” and “Pervasive Intelligence” - but in recent years, it has gained critical mass. In order to realize the vision of a worldwide network of smart objects, paradigms from several fields must be understood and unified - protocols and middleware, embedded design, agent-based systems, human-computer interaction, and so on. This requires a common vocabulary. Also, without a way to model assets, adversaries, and threat analysis, our understanding is too limited to provide safety guarantees (as seen in the Stuxnet worm, Duqu, etc.); partial solutions, like ports of IPSec and SSL to sensor networks\cite{hummen2013}, are not enough, especially when we consider the high cost of failure in physical systems.

In this paper, we build upon the concepts of Object-Oriented and Stratified Design to propose a general vocabulary and model for the Internet of Things. Our principal idea: the cell, a unit within which all component things share the same context. We hope to develop this model into a theory of cells, with operations such as union and intersection of cells of different types etc., and propose a formal model for security, in our future work.

\bibliographystyle{abbrv}
\bibliography{min}

\begin{thebibliography}{1}

\bibitem{abowd1999}
G.~D. Abowd, A.~K. Dey, P.~J. Brown, N.~Davies, M.~Smith, and P.~Steggles.
\newblock Towards a better understanding of context and context-awareness.
\newblock In {\em Handheld and ubiquitous computing}, pages 304--307. Springer,
  1999.

\bibitem{atzori2010}
L.~Atzori, A.~Iera, and G.~Morabito.
\newblock The internet of things: A survey.
\newblock {\em Computer networks}, 54(15):2787--2805, 2010.

\bibitem{burrus2014}
D.~Burrus.
\newblock The internet of things is far bigger than anyone realizes.
\newblock 2014.

\bibitem{hewitt1978}
C.~Hewitt and H.~G. Baker.
\newblock Actors and continuous functionals.
\newblock Technical report, Cambridge, MA, USA, 1978.

\bibitem{hummen2013}
R.~Hummen, J.~H. Ziegeldorf, H.~Shafagh, S.~Raza, and K.~Wehrle.
\newblock Towards viable certificate-based authentication for the internet of
  things.
\newblock In {\em Proceedings of the 2nd ACM workshop on Hot topics on wireless
  network security and privacy}, pages 37--42. ACM, 2013.

\bibitem{kay1977}
A.~Kay and A.~Goldberg.
\newblock Personal dynamic media.
\newblock {\em Computer}, 10(3):31--41, Mar. 1977.

\bibitem{kelly2007}
K.~Kelly.
\newblock Four stages in the internet of things, 2007.

\bibitem{neisse2014}
R.~Neisse, I.~N. Fovino, G.~Baldini, V.~Stavroulaki, P.~Vlacheas, and
  R.~Giaffreda.
\newblock A model-based security toolkit for the internet of things.
\newblock In {\em Availability, Reliability and Security (ARES), 2014 Ninth
  International Conference on}, pages 78--87. IEEE, 2014.

\bibitem{roman2013}
R.~Roman, J.~Zhou, and J.~Lopez.
\newblock On the features and challenges of security and privacy in distributed
  internet of things.
\newblock {\em Computer Networks}, 57(10):2266--2279, 2013.

\end{thebibliography}

\end{document}